\documentclass{sigchi}
\CopyrightYear{2020}
\setcopyright{acmlicensed}
\doi{https://doi.org/10.1145/3313831.XXXXXXX}
\isbn{978-1-4503-6708-0/20/04}
\conferenceinfo{CHI'20,}{April  25--30, 2020, Honolulu, HI, USA}
\acmPrice{\$15.00}

\usepackage{balance}       
\usepackage{graphics}      
\usepackage[T1]{fontenc}   
\usepackage{txfonts}
\usepackage{mathptmx}
\usepackage[pdflang={en-US},pdftex]{hyperref}
\usepackage{color}
\usepackage{booktabs}
\usepackage{textcomp}

\usepackage{microtype}        
\usepackage{ccicons}          

\usepackage{todonotes}

\def\plaintitle{Data Model, Collection and Evaluation Framework for Local Energy Systems}
\def\plainauthor{First Author}

\def\plainkeywords{data models, distributed systems, de-centralized, blockchain, consensus, data stores, data integrity, relations, data access, services}

\makeatletter
\def\url@leostyle{%
  \@ifundefined{selectfont}{
    \def\UrlFont{\sf}
  }{
    \def\UrlFont{\small\bf\ttfamily}
  }}
\makeatother
\def\pprw{8.5in}
\def\pprh{11in}

\definecolor{linkColor}{RGB}{6,125,233}
\hypersetup{%
  pdftitle={\plaintitle},
  pdfauthor={\plainauthor},
  pdfkeywords={\plainkeywords},
  pdfdisplaydoctitle=true, 
  bookmarksnumbered,
  pdfstartview={FitH},
  colorlinks,
  citecolor=black,
  filecolor=black,
  linkcolor=black,
  urlcolor=linkColor,
  breaklinks=true,
  hypertexnames=false
}
\begin{document}
\title{\plaintitle}
\numberofauthors{1}
\author{%
  \alignauthor{Chitti S Phani\\
    \affaddr{University of Bristol}\\
    \affaddr{Bristol, UK}\\
    \email{sc18092@bristol.ac.uk}}\\
}

\maketitle

\begin{abstract}
Distributed ledgers are a new type of database technology that allows open access to data stored across distributed, decentralised, publicly maintained infrastructures. Current implementations of the such ledgers expect competition between participants, are often energy hungry, poor in maintaining the natural structure of data and suffer from scalability constraints. The aim of my research work is to develop a distributed ledger-based middleware for data modelling and collection on household energy generation and use, while addressing scalability and energy inefficiency concerns of the ledger for this particular application domain. The energy data collected and made available through this middleware will be used for digital energy service delivery (e.g., automated peer to peer energy trading, topological estimations, etc.). The middleware also provides a platform for a consumer focused digital energy service delivery, as well as service model evaluation. The model evaluation will enable the prospective service users to evaluate the suitability of the given service for their needs before making a decision of service subscription.
\end{abstract}



\section{Context and Motivation}
The term Database has been defined by  CJDate \cite{10.5555/940536} as "a collection of related data stored on a computer that can be used for different applications without knowledge of storage details". Modelling data is the central characteristic that any database (DB) should have, such that the management system (MS) around it can free it\textquoteright s users from machine representation and organization of data. DBMS implementations based on relational model proposed by EF Codd \cite{Codd2009DerivabilityRA} uses role based management system, controlled centrally by a group of administrators, to make the state changing in data efficient and structured.\\
This leads to an interesting question  "are centralized and role based DBMSs efficient in storing data that is immutable in their life cycle?". Pat Helland defined the immutability  \cite{10.1145/2857274.2884038} of data by categorizing data as "Data on the inside" and "Data on the outside". The data sent out from a service and/or web page is in the second category and such data is immutable, semi-structured, has identity and may be versioned. The updates to such data creates a new version data with a new unique identifier. A Dataset is a collection of such data with a unique ID. In other words datasets are immutable collection of data. The data collected andß bundled as datasets from local energy systems is either from tailor made services hosted in servers or from internet-of-things (IoT) devices.\\
After defining immutable data, the above question can be modified as "can the datasets that are given by local energy systems be stored efficiently using centralized databases?". Finding an answer to this question is the motivation for my current research work.\\
The datasets that are emerging from local energy systems contain discrete sets of time-series data values e.g., indicating sensor readings or any other information. The distinctive characteristic of these datasets is that they are cohesive reflecting the natural structure of underlying system. Keeping the cohesion among these discrete datasets is important for service evaluation and modelling tasks.\\
Traditionally the centralized databases are rather best fit to store the data that belongs to first category, "Data on the inside", because such data lives in transactional world \cite{10.1145/2857274.2884038}. Another possible way to store the datasets, could be using NoSQL databases, which store data in a schema-less fashion \cite{nosql}. The popular No-SQL models are: 1) Document databases (e.g., Mongo DB) where data stored as free form JSON structure; 2) Key-Value stores (e.g., Redis) values  like strings or JOSN are accessed using keys; 3) Wide column stores(e.g., Casandra) where data is stored in columns; 4) Graph database (e.g., Neo4J) in which the data represented as graph of nodes and relationships.\\ 
It appears that the Key-Value and Graph databases are not suitable to store the datasets, because of the data model these databases offer. Storing discrete data in the free form or in columns stores may serve as half of a solution. These, however, do not address the second half of the problem, that is preserving natural cohesion among stored datasets. Both types of databases: relational and No-SQL, lack such features (sought after features for immutable collection of data) as open access and transparency.\\
Distributed ledger technologies (a.k.a. blockchains) are gaining more recognition as decentralized data stores and as an alternative to the above mentioned data storage systems. Initially these technologies arose to support crypto currencies \cite{nakamoto2008bitcoin}. However, their distinguishing features  -  distributed consensus, non-dependency on a centralized authority, built-in trust, open and low entry barriers for joining the network, transparency in maintaining the ledger - are attracting many diverse domains to adapt this technology. However the limitations of the blockchain technology  are it\textquoteright s inability to scale as per demand, and high read and write latency \cite{10.1007/978-3-030-20948-3_17}. Also the blockchain does\textquoteright t have inherent capability to preserve the natural cohesion among datasets \cite{10.1007/978-3-030-21297-1_5}.\\
The \textbf{aim} of my research work is to \textit{develop a distributed ledger-based middleware, for modelling and collecting datasets that arises from local energy systems, while addressing scalability and energy inefficiency concerns of the ledger.}
The data to be stored in such a database would predominantly be of immutable nature and can come from different application domains like local energy systems, lighting and electric heating data, power distribution networks with distributed energy resources(DER) or even from data banks like UK Biobank \cite{biobank} or  Genetic database \cite{gendata}. \\
The Research Questions of this work are  presented in following section. These are followed by the summary of the conducted literature review; the proposed approach for solution; research progress so far;  and conclusions. 
\section{Research Questions}\label{rq}
The below research questions (RQ) anchor the various research tasks: from  literature review, to feasible solution elibroration:  
\begin{itemize}
    \item RQ1: How is blockchain technology adapted in various domains?\\
            This question aims to elicit how different application areas apply blockchain for  data storage tasks, and are used by software applications and as an alternative to traditional databases.  
    \item RQ2: What data is stored in a blockchain? 
        \begin{itemize}
            \item Structured or unstructured (e.g., images, documents, logs etc).
            \item Discrete or related
        \end{itemize}
    \item RQ3: How are the below data quality features realized?
        \begin{itemize}
            \item Data integrity: i.e.,  completeness, accuracy and consistency of data;
            \item Data Access: obtaining required data from underlying data store;
            \item Data Indexing: optimizing performance of the underlying data store by minimizing the number of disk accesses required;
            \item Physical Data Storage: the way that data is physically stored on a disk.
        \end{itemize}
    \item RQ4: How are the below distributed system characteristics of \textit{Trusted/un-trusted networks}, \textit{Synchronized/non-synchronized network} (clock synchronization), \textit{Data replication strategy} used?
        In short, we view blockchains as a kind of distributed system. Thus we think that mapping the current practice of its use against theoretical concepts will provide relevant insights.  
    \item RQ5: How are the (business-related) features of \textit{Scalability}, \textit{Consisstency}, and \textit{Read/Write latency} used and modified?
\end{itemize}

\section{Approach}\label{aproach}
To address the above questions, a number of pre-development tasks are identified,  such as:
\begin{itemize}
    \item Conduct \textit{Literature review} in order to understand the current progress in similar research areas, and to obtain the technical details that drive the existing blockchain solutions like Hyperledger \cite{Androulaki:2018:HFD:3190508.3190538}, IOTA \cite{iota:2019:Online}, BigchainDB \cite{bdb:2019:Online}, Tangle \cite{Popov2015TheT}. 
    \item \textit{Review energy sector datasets} in order to get a proper insight into datasets from different application domains. It is hoped that a comprehensive knowledge about data life cycle in datasets from different domains will be observed and knowledge  about the natural cohesion among such datasets  will be gained. 
    \item Gaining perspective on \textit{Technical details}, given that blockchains are a kind of distributed system,  getting theoretical perspective on issues within the existing distributed systems (like network consensus, redundancy, state changing, etc.) will be informative. Similarly, knowledge of  existing distributed databases (e.g., Spanner \cite{10.1145/2491245}, Bigtable \cite{10.1145/1365815.1365816} ) and file systems (e.g., Google File System \cite{10.1145/945445.945450}, IPFS \cite{ipfs}) will be equally relevant.  
    \item \textit{Identify viable technical solutions}, whereby suitable technical solutions for crucial components of the solution would be identified. The crucial components are 
    \begin{itemize}
        \item Network Creation and participation
        \item Consensus among participants
        \item State changing
        \item Data storage
        \item Redundancy of data in network
        \item Data indexing and access
    \end{itemize}
\end{itemize}
Once these tasks are completed, the software development and testing cycle will be iterated. However, the development tasks are presently not detailed upon.

\section{Progress so far}\label{lit}
Below we briefly outline the work carried out so far.
\subsection{Literature Review}\label{lit}
A literature review on use of blockchains as a data storage platform is carried out. For this a systematic literature review protocol (proposed by Kitchenham \cite{Kitchenham:2009:SLR:1465742.1466091}) is used. The review protocol defines: 1) Search Strategy; 2) Study Selection; 3) Data Extraction and 4) Data Synthesis tasks.
As part of the Search Strategy, relevant search terms (namely: \textit{Blockchain, Data Storage, Ledger, Data Integrity}) were identified.  These terms were formed into search strings: \textit{Blockchain AND (Data Store OR Data storage) AND (Transparent OR Open OR Decentralized OR Distributed OR Immutable AND Ledger} and \textit{Data integrity OR Data quality OR Data consistency OR Related data OR Referential Integrity}. The  IEEE and Bristol Library Search repositories were used for the document search, from which 185 relevant papers were identified.\\
The inclusion criteria specifying the publications of type (i.e., peer reviewed conference and workshop papers) and content (papers that cover the architectural details of software that use  blockchain to store data and uses  distributed system platform features for data management) was used. On basis these criteria, 113 papers were chosen for detailed review from the initial corpus of 185.\\
For Data Extraction, the search parameters based on features from blockchain and distributed systems along with data characteristics were used. The extraction results are detailed in  \cite{dataextraction}.\\ 
The brief summary of the Data Synthesis suggests the below key findings:
\begin{itemize}
    \item Applicability: The details of application areas that use blockchain to store different kinds of data are extracted and synthesized. The Fig - \ref{fig:aa-dist} shows the types of data stored in blockchain by different applications.
    \begin{figure}[h]
    \centering
    \includegraphics[width=\linewidth]{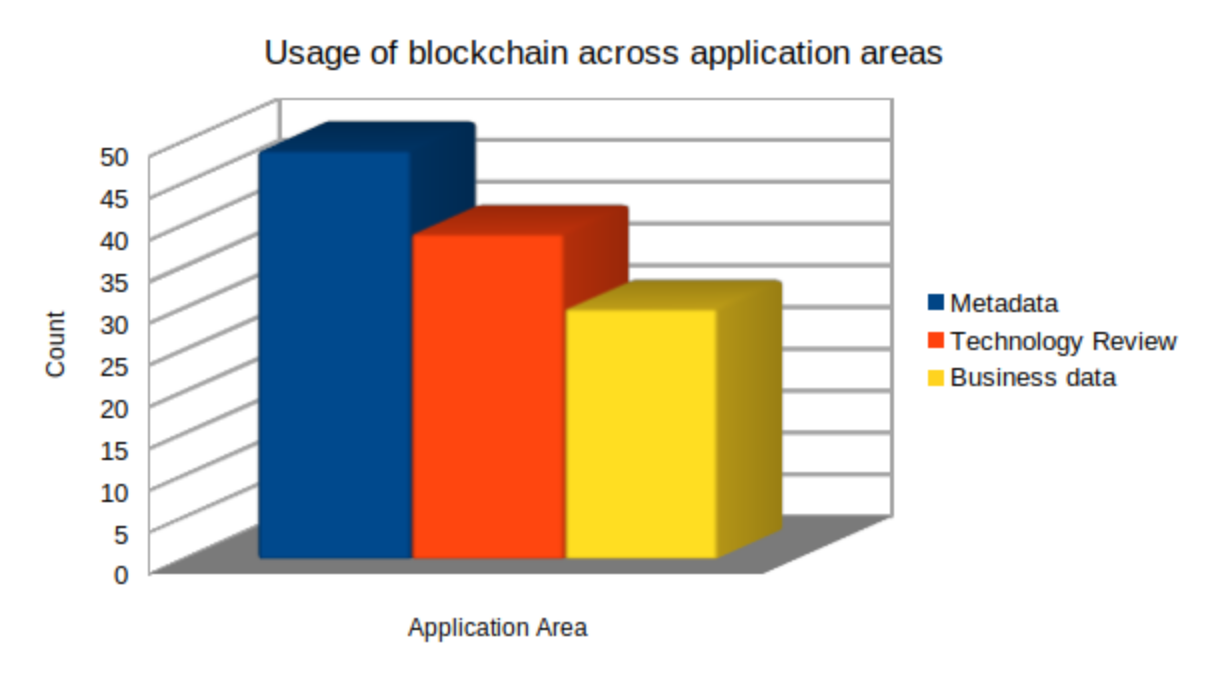}
    \caption{Usage of blockchain across application areas}
    \label{fig:aa-dist}
    \end{figure}
    As shown in the Fig-\ref{fig:aa-dist}, majority of applications use blockchain to  store  metadata  that summarizes  underlying business  data, to  ensure  the  integrity  of such data, and to provide trust factor to customers.
    \begin{figure}[h]
    \centering
    \includegraphics[width=\linewidth]{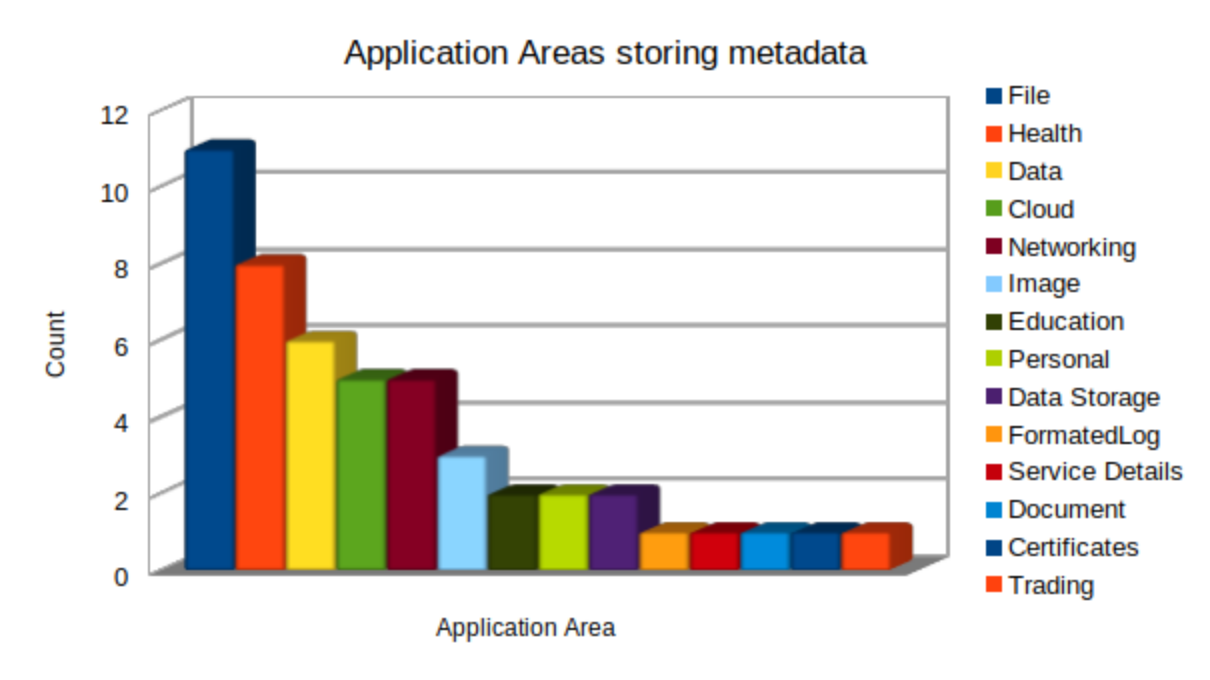}
    \caption{Application areas using blockchain to store meta data}
    \label{fig:aa-metadata-dist}
    \end{figure} 
    The Fig - \ref{fig:aa-metadata-dist} shows those application areas, which use blockchain for storing meta data. However, considerable application areas store data emerging from them in blockchain as shows in Fig - \ref{fig:aa-bdata-dist}
    \begin{figure}[h]
    \centering
    \includegraphics[width=\linewidth]{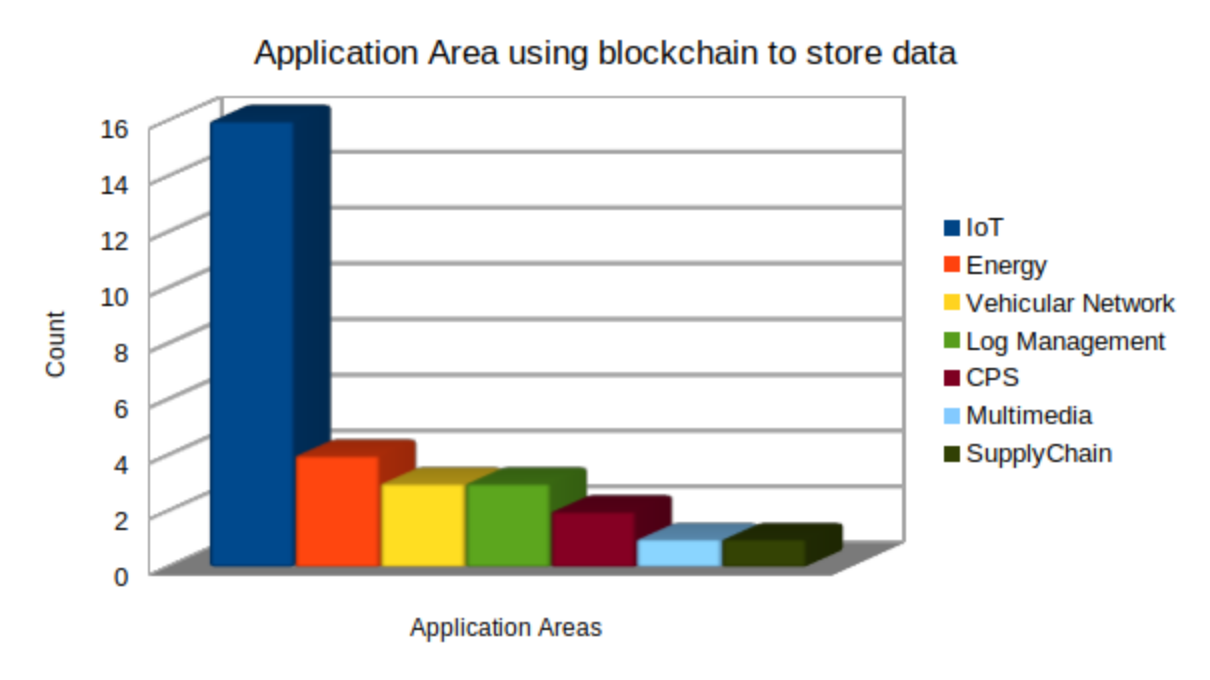}
    \caption{Application areas using blockchain to store data}
    \label{fig:aa-bdata-dist}
    \end{figure} 
    \\The key finding from the above data synthesis results is blockchain is used to store datasets. Both metadata and the data emerging from applications are immutable and applications prefer blockchain to store such data to ensure integrity of data and to provide the trust factor to customers.  
    \item Blockchain Variants and networks: The details of blockchain variants used by software designers to store data in blockchain are extracted and analyzed. 
    \begin{figure}[h]
    \centering
    \includegraphics[width=\linewidth]{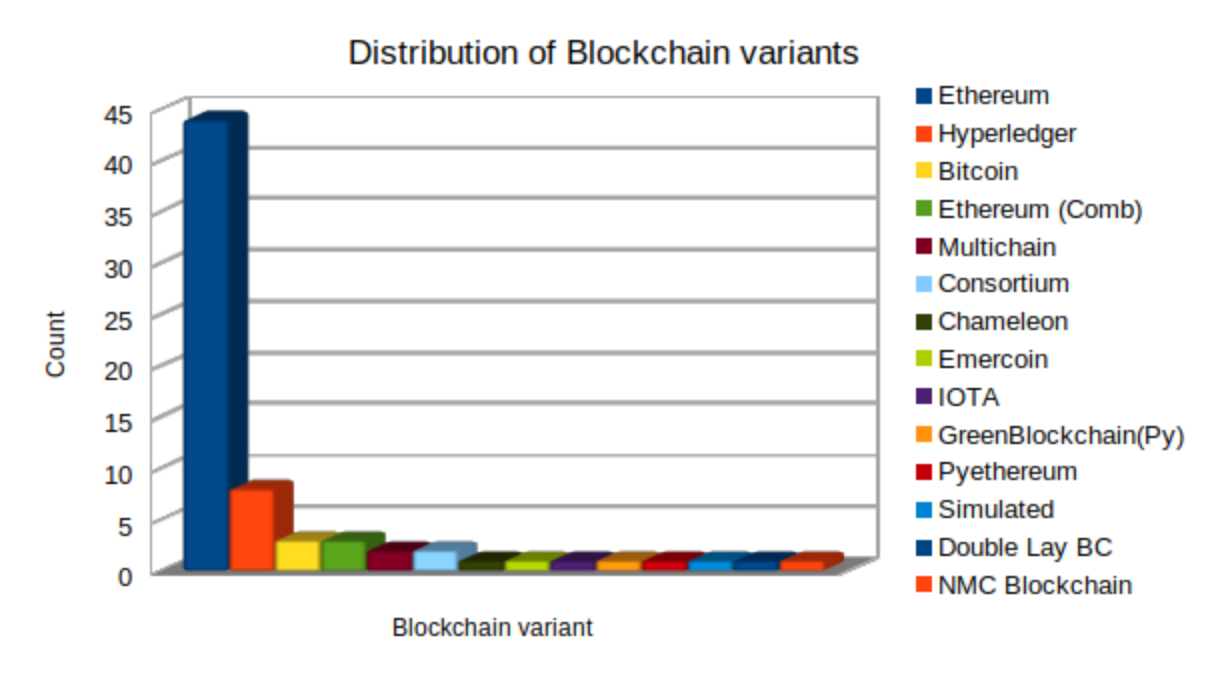}
    \caption{Distribution of blockchain variants across application areas}
    \label{fig:bcvariantdist}
    \end{figure}
    The Fig - \ref{fig:bcvariantdist} shows the choice of blockchain variants
    \\Ethereum \cite{ethsc:2019:Online} is the most used blockchain variant followed by Hyperledger \cite{Androulaki:2018:HFD:3190508.3190538}. Ability to write scripts to customize the state changing in Ethereum blockchain is the main motivation factor. Hyperledger provide more flexibility in terms of customization and getting more prominence. 
    \\Another important factor in using blockchain is the type of it\textquoteright s network. Three types of Blockchain networks: private, public consortium or permissioned are most used as shown in Fig - \ref{fig:nwtypedist}.
    \begin{figure}[h]
    \centering
    \includegraphics[width=\linewidth]{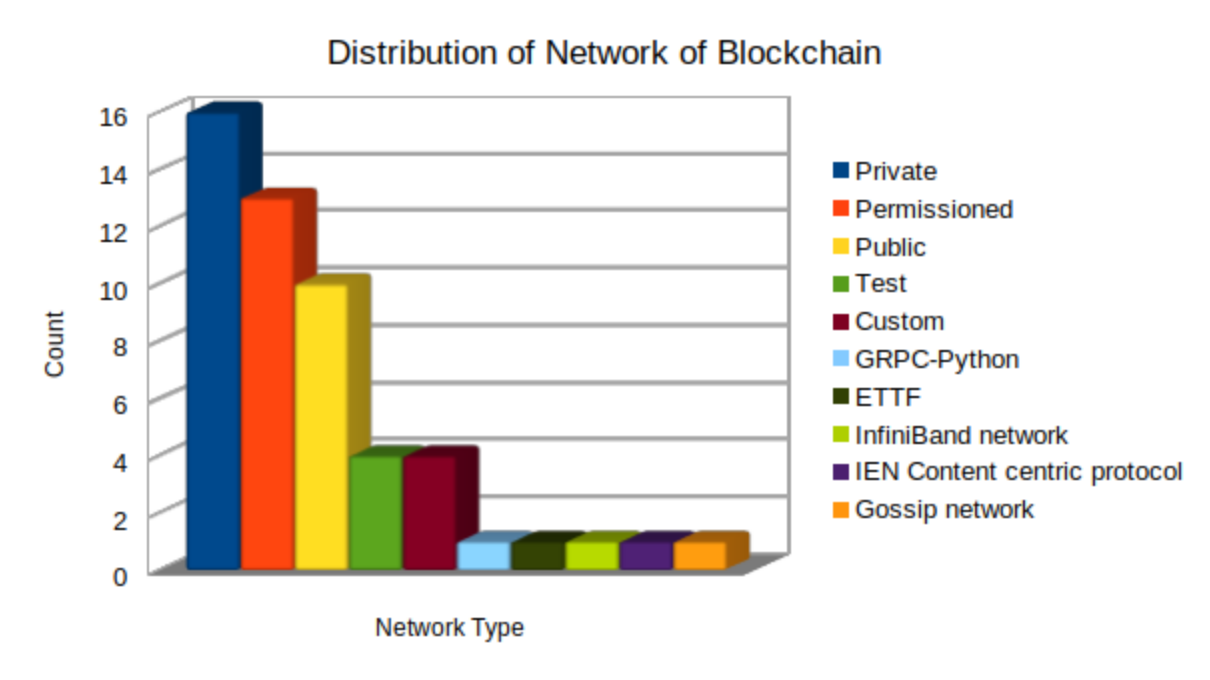}
    \caption{Distribution of blockchain networks across application areas}
    \label{fig:nwtypedist}
    \end{figure}
    Blockchain with private networks is the popular choice for software designers. The motivation is to avoid the GAS charges involved and to reduce the read/write latency. Consortium/permissioned networks are next popular choice, which are created and maintained by group of companies such that network management is controlled and participation is not open.  \\
    A key finding from the above analysis is that choice of a blockchain variant and network is dependent on the data to be stored and the data access requirements. 
\end{itemize}

\subsection{Review energy sector datasets}\label{experiments}
To gain a better understanding of the blockchain technology, the Ethereum blockchain \cite{ethsc:2019:Online} was used to prototype a solution to to store sample energy datasets. The datasets related to \textit{peer-to-peer (P2P) energy trading}  and\textit{ occupancy analysis within the university accommodations}. The prototyping effort focused on understanding the process of data modelling in Ethereum and maintaining relations between different datasets (using smart contracts). The findings are presented in \cite{10.1007/978-3-030-20948-3_17,10.1007/978-3-030-21297-1_5}.
 
To further our perspectives on the nature of the data from different application domains, and to expriment with different implementations with data models,  data sets from \textit{low voltage distribution stations} (OpenLV) and  university's \textit{lighting system} were studied. 
The OpenLV \cite{openlv} data was obtained from sensors deployed at different components of substations. The data was modelled using relational data model and hosted in a relationoal database. The size and data access times for data hosted in RDBMS were recorded. At the same time, the lighting data was modelled as smart contracts. A middleware prototype was created using private Ethereum blockchain (GETH) such that data can be stores and accessed.

Comparative analysis of these data storage and modelling solution prototypes are presently under our review. 

\section{Future Work}\label{conclusions}
In the first year of PhD the preliminary work was done to understand the landscape of the research by means of experimentation and studying literature. 

Next we will focus on technical perspective for middleware along with identifying the efficient technical solutions, as discussed above.
\section{Acknowledgments}
I thank my mentor Dr. R Chitchyan, and collegue J. Murkin for introducing me to the P2P trading using blockchain.
\balance{}
\balance{}
\bibliographystyle{abbrv}
\bibliography{sample}

\end{document}